\documentclass[aps,prb,twocolumn,superscriptaddress,nofootinbib,showpacs]{revtex4}
\usepackage{amssymb}
\usepackage{graphicx}
\usepackage{epsf,epsfig}
\usepackage{bm}
\usepackage{bbm}
\usepackage{amsfonts}
\usepackage{amsmath}
\usepackage{graphics}
\usepackage[normalem]{ulem}
\usepackage{color}

\newcommand{\be}{\begin{eqnarray}}
\newcommand{\ee}{\end{eqnarray}}
\newcommand{\ba}{\begin{array}}
\newcommand{\ea}{\end{array}}
\newcommand{\no}{\nonumber}

\newcommand{\Tr}{\mbox{Tr}}

\newcommand{\eps}{\varepsilon}

\newcommand{\bfr}{{\bf r}}

\newcommand{\bfk}{{\bf k}}

\newcommand{\bfq}{{\bf q}}
\newcommand{\bfs}{{\bf s}}

\newcommand{\bfsigma}{{\bm \sigma}}

\begin{document}

\title{ Theory of Thermal Conductivity in the Disordered Electron Liquid}
\author{G. Schwiete}
\email{schwiete@uni-mainz.de}
 \affiliation{Spin Phenomena Interdisciplinary Center (SPICE) and Institut f\"ur Physik,
Johannes Gutenberg Universit\"at Mainz, 55128 Mainz, Germany}
\author{A. M. Finkel'stein}
\affiliation{Department of Physics and Astronomy, Texas A\&M University, College Station, TX 77843-4242, USA}
\affiliation{Department of
Condensed Matter Physics, The Weizmann Institute of Science, 76100
Rehovot, Israel}
\affiliation{L. D. Landau Institute for Theoretical Physics, 117940 Moscow, Russia}
\date{\today}

\begin{abstract}
We study thermal conductivity in the disordered two-dimensional electron liquid in the presence of long-range Coulomb interactions. We describe a microscopic analysis of the problem using as a starting point the partition function defined on the Keldysh contour. We extend the Renormalization Group (RG) analysis developed for thermal transport in the disordered Fermi liquid, and include scattering processes induced by the long-range Coulomb interaction in the sub-temperature energy range. For the thermal conductivity, unlike for the electric conductivity, these scattering processes yield a logarithmic correction which may compete with the RG-corrections. The interest in this correction arises from the fact that it violates the Wiedemann-Franz law. We checked that the sub-temperature correction to the thermal conductivity is not modified, neither by the inclusion of Fermi
liquid interaction amplitudes nor as a result of the RG flow. We thereby expect that the answer obtained for this correction is final. We use the theory to describe thermal transport on the metallic side of the metal-insulator transition in Si MOSFETs.

\end{abstract}

\pacs{71.10.Ay, 72.10.-d, 72.15.Eb, 73.23.-b} \maketitle

\section{Introduction}

At temperatures smaller than the impurity scattering rate, i.e., in the diffusive regime, the electron liquid acquires various non-analytic quantum corrections [\onlinecite{Altshuler85, Finkelstein90}]. At low temperatures, the calculation of these corrections demands an RG analysis which leads to coupled flow equations for the diffusion constant, the interaction constants, and also the frequency coefficient [\onlinecite{Finkelstein83,Castellani84,Finkelstein84,Baranov99, Schwiete14}]; for a review see [\onlinecite{Finkelstein90,Belitz94RMP,DiCastro04,Finkelstein10}]. A systematic procedure for the derivation of the RG-equations in disordered electron systems has been developed on the basis of a field-theoretic description [\onlinecite{Finkelstein83}], the nonlinear sigma model (NL$\sigma$M). In a series of recent papers [\onlinecite{Schwiete14a,Schwiete14b, Schwiete15}], we extended the NL$\sigma$M formalism for the study of thermal transport. To this end, we introduced time-dependent "gravitational potentials" [\onlinecite{Luttinger64,Shastry09, Michaeli09}] as source fields into the microscopic action. These sources are a convenient tool for generating expressions for the heat density-heat density correlation function from the partition function, which is defined on the Keldysh contour. Knowledge of the correlation function then allows to obtain the thermal conductivity.

In this paper, we include both the long-range Coulomb interaction and Fermi-liquid type interactions. Besides the RG-corrections, we will focus our attention on the sub-temperature energy range, which is beyond the scope of the conventional RG analysis. The main difference between the RG-interval $(1/\tau, T)$ and the sub-temperature energy range is that, while the transitions described by the standard RG procedure are virtual, the sub-thermal range deals with \emph{on-shell} scattering. For the analysis of logarithmic corrections to electric conductivity, the sub-thermal processes can be neglected. For thermal conductivity, however, the sub-thermal scattering processes induced by the long-range Coulomb interaction also yield logarithmic corrections. These corrections violate the Wiedemann-Franz law (WFL) [\onlinecite{Wiedemann1853}]. The calculation of thermal conductivity therefore demands a two-stage procedure: the leading logarithmic corrections originating from energies in the RG-interval $(1/\tau,T)$ can be absorbed into the scale-dependent RG charges of the \emph{extended} NL$\sigma$M, i.e., the model which also includes the gravitational potentials. Once all RG-corrections are taken into account, one may find the sub-temperature correction using parameters determined by the current scale of the RG procedure. As we will show, however, the sub-temperature corrections to the thermal conductivity caused by the long-range Coulomb interaction remain largely unaffected by the renormalizations. With this result at hand, we can apply the theory to the metallic side of the metal-insulator transition in Si MOSFETs.

The paper is organized as follows: In Sec.~\ref{sec:KeldyshAction} we introduce the Keldysh action and collect basic formulas for the calculation of the heat density-heat density correlation function. Sec.~\ref{sec:GravPot} is devoted to the discussion of the extended NL$\sigma$M in the presence of gravitational potentials. In Sec.~\ref{sec:RGanalysis}, we obtain the RG equations describing the flow of the gravitational potentials as a function of temperature. Secs.~\ref{sec:RGanalysis2} and \ref{sec:sub} deal with quantum corrections to the heat density-heat density correlation function. First, in Sec.~\ref{sec:RGanalysis2}, we include the corrections originating from the RG-interval of energies, and subsequently, we discuss sub-thermal corrections in Sec.~\ref{sec:sub}. The developed theory is applied to the description of thermal transport on the metallic side of the metal-insulator transition in Si MOSFETs in Sec.~\ref{sec:MOSFET}. The results for the thermal resistance are illustrated in Fig.~\ref{fig:Rkofeta}.

\section{Keldysh action and the correlation function}
\label{sec:KeldyshAction}
Here, we follow the approach developed by us in Refs. [\onlinecite{Schwiete14a,Schwiete14b, Schwiete15}]. We introduce the Keldysh partition function as $\mathcal{Z}=\int D[\psi^\dagger,\psi] \exp(iS[\psi^\dagger,\psi])$.  The action is first limited to $S=S_k$, where
\be
S_{k}[\psi^\dagger,\psi]=\int_\mathcal{C}dt \int_{\bfr} \left(\psi^\dagger i \partial_t\psi-k[\psi^\dagger,\psi]\right)
\ee
is defined on the Keldysh contour $\mathcal{C}$ [\onlinecite{Keldish65,Kamenev11}]. Here, $k=h-\mu n$, where $h$ is the hamiltonian density, $n$ is the particle density, $\mu$ is the chemical potential, and $\psi=(\psi_\uparrow,\psi_\downarrow)$, $\psi^\dagger=(\psi^*_\uparrow,\psi^*_{\downarrow})$ are vectors of Grassmann fields accounting for the electrons with two spin components. Interestingly, $S_k$ is mainly determined by the heat density $k$ [\onlinecite{Heat}], i.e., by the quantity we study.

The retarded heat density-heat density correlation function is defined as
\be
\chi_{kk}(x_1,x_2)=-i\theta(t_1-t_2)\langle[\hat{k}(x_1),\hat{k}(x_2)]\rangle_T,
\ee
where $x=(\bfr,t)$ and $\hat{k}=\hat{h}-\mu\hat{n}$ is the heat density operator. The angular brackets denote thermal averaging. After introducing fields on the forward ($+$) and backward ($-$) paths of the Keldysh contour, one may define classical ($cl$) and quantum components ($q$) of the heat density, $k_{cl/q}=\frac{1}{2}(k_+\pm k_-)$ [\onlinecite{Kamenev11}]. Then, the retarded correlation function can be obtained as $\chi_{kk}(x_1,x_2)=-2 i\left\langle k_{cl}(x_1) k_q(x_2)\right\rangle$. Here, the averaging is with respect to the action $S_k$. This representation of the correlation function motivates us to introduce the source term
\be
S_\eta=2\int_x [\eta_2(x)k_{cl}(x)+\eta_1(x)k_q(x)]
\ee
into the action $S=S_{k}+S_\eta$, which allows to generate $\chi_{kk}$ as
\be
\chi_{kk}(x_1,x_2)&=&\frac{i}{2}\left.\frac{\delta^2 \mathcal{Z}}{\delta \eta_2(x_1)\delta \eta_1(x_2)}\right|_{\eta_1=\eta_2=0}.
\ee

The thermal conductivity $\kappa$ is related to the disorder-averaged correlation function $\langle\chi_{kk}(x_1,x_2)\rangle_{dis}=\chi_{kk}(x_{1}-x_2)$  as [\onlinecite{Castellani87}]
\be
\kappa =-\frac{1}{T}\lim_{\omega\rightarrow 0}\left(\lim_{q\rightarrow 0}\left[\frac{\omega}{\bfq^2} \mbox{Im}\chi_{kk}(\bfq,\omega)\right]\right)\label{eq:continuity}.
\ee
This expression is typical for a transport coefficient related to a conserved quantity:
We study heat conductivity in a situation where mechanical work (e.g., caused by thermal expansion, or radiation of acoustic waves) can be neglected.

\section{Gravitational potentials and NL${\bm \sigma}$M}
\label{sec:GravPot}
The hamiltonian density $h=h_0+h_{int}$ is chosen to describe the electron liquid in a static disorder potential.
The non-interacting part of the Hamiltonian density is
\be
h_{0}&=&\frac{1}{2m^*}\sum_{\alpha}\nabla\psi^*_{\alpha}(x)\nabla{\psi}_{\alpha}(x)+u_{dis}(\bfr)n(x),
\ee
where $u_{dis}(\bfr)$ is the disorder potential and $m^*$ is the (renormalized) mass.
The interaction will be subdivided into singlet and triplet parts
\be
h_{int,\rho}&=&\frac{1}{2}\int_{\bfr'}\; n(\bfr,t)\;V_{\rho}(\bfr-\bfr')\;n(\bfr',t)\label{eq:vrho}\\
h_{int,\sigma}&=&2 \int_{\bfr'} \;\bfs(\bfr,t)\;V_{\sigma}(\bfr-\bfr')\;\bfs(\bfr',t),
\ee
where we introduced the particle-number density and spin densities
\be
n(x)={\psi}^\dagger_x\sigma^0\psi_x,\qquad \bfs(x)=\frac{1}{2}{\psi}^\dagger_x \bfsigma \psi_x .
\ee
The interactions in the singlet and triplet channels are described by the amplitudes
\be
V_{\rho}(\bfq)= V_{0}(\bfq)+\frac{F_0^\rho}{2\nu_0}, \quad V_{\sigma}(\bfq)=\frac{F_0^\sigma}{2\nu_0}.
\ee
Here, $F_0^\rho$ and $F_0^\sigma$ are the phenomenological Fermi liquid parameters [\onlinecite{AGD63,Lifshitz80}], and $\nu_0$ is the single-particle density of states per spin direction. In $V_{\rho}(\bfq)$, the bare long-range part of the Coulomb interaction $V_0 (\bfq)$ is separated from the short-range part; the $q$-dependence of $V_{\sigma}(\bfq)$ is redundant. We anticipate that in the diffusive limit, $T\tau\ll 1$, which we will study here, only the zeroth angular harmonics will be effective.

To proceed further, we perform the Keldysh rotation [\onlinecite{Larkin75,Kamenev11}], and decouple the interaction terms using a Hubbard-Stratonovich field $\theta_k^l$, where the index $k=1,2$ counts the two Keldysh components ($1,2$ correspond to $cl,q$), and the index $l=0-3$ denotes the density and spin density interaction channels. After this decoupling one can write the action as
\begin{align}
{S}=&\int_x \;\Psi^\dagger \{i\partial_t-[u_{dis}-\mu](1+\hat{\eta})+\hat{\theta}^l\sigma^l\}\Psi \no\\
&-\int_x\frac{1}{2m^*}\nabla \Psi^\dagger(1+\hat{\eta})\nabla \Psi +\int_x \vec{\theta}^T\frac{\hat{\gamma}_2}{1+\hat{\eta}}V^{-1}\vec{\theta}.\label{eq:S3}
\end{align}
From now on, $\Psi(x)$ and $\Psi^\dagger(x)$ are fields with two Keldysh components (their spin indices are not displayed); the hat symbol indicates matrices in Keldysh space. The matrices $\hat{\theta}$ and $\hat{\eta}$ are defined as $\hat{\eta}=\sum_{k=1,2}\eta_k\hat{\gamma}_k$, $\hat{\theta}^l=\sum_{k=1,2}\theta^l_{k}\hat{\gamma}_k$, where $\hat{\gamma}_1=\hat{\sigma}_0$, $\hat{\gamma}_2=\hat{\sigma}_x$ and $\hat{\sigma}_0$, $\hat{\sigma}_x$ are Pauli matrices in Keldysh space. The Pauli matrices $\sigma_l$ in Eq.~\eqref{eq:S3} act in spin space. The matrix $V=\mbox{diag}[V_\rho,V_\sigma,V_\sigma,V_\sigma]$ distinguishes the different interaction channels. An \emph{important remark} is in order here. In the presentation of  Eqs.~\eqref{eq:vrho}, \eqref{eq:S3} we ignored subtle questions related to the procedure of introducing gravitational potentials for the long-range potential $V_0(\bfq)$. A detailed discussion of these questions can be found in Ref.~\onlinecite{Schwiete15}.

A disadvantage of the representation in Eq.~\eqref{eq:S3} is that the gravitational potentials couple directly to the disorder potential $u_{dis}$, a fact that complicates the derivation of the NL$\sigma$M. As discussed in Refs.~\onlinecite{Schwiete14a,Schwiete14b,Schwiete15}, one can perform a transformation of the fields $\psi$ and $\bar{\psi}$ that removes the gravitational potential from the disorder term. The mentioned transformation reads as $\psi\rightarrow \sqrt{\hat{\lambda}}\psi$, $\bar{\psi}\rightarrow \bar{\psi} \sqrt{\hat{\lambda}}$, where $ \hat{\lambda}=1/(1+\hat{\eta})$. A term proportional to $(\nabla\hat{\eta})^2$ emerging from this transformation may be ignored due to the slowness of the gravitational potentials acting as source fields. [The Jacobian arising from the $\lambda$-transformation is featureless [\onlinecite{Schwiete14b}].] As a result, the gravitational potentials are removed from $h_{0}-\mu n$. At the same time, they reappear in the time-derivative term and also change the structure of the interaction part
\be
S&=&\frac{1}{2}\int_x {\Psi}^\dagger(i\hat{\lambda}\overrightarrow{\partial}_t-i\overleftarrow{\partial}_t\hat{\lambda})\Psi\no\\
&&-\int_x \;\Psi^\dagger (u_{dis}-\mu-\hat{\lambda}\hat{\theta}^l\sigma^l)\Psi\no\\
&&-\int_x\frac{1}{2m^*}\nabla \Psi^\dagger\nabla\Psi+\int_x \vec{\theta}^T(\hat{\gamma}_2\hat{\lambda})V^{-1}\vec{\theta}.
\label{eq:Sintermediate}
\ee
Starting from Eq.~\eqref{eq:Sintermediate}, the NL$\sigma$M can be derived according to the standard scheme [\onlinecite{Wegner79,Efetov80,Finkelstein90, Kamenev99, Chamon99,Schwiete14}]; the part of the action that describes diffusion modes may be written in the form
\be
S_{dm}&=&\frac{\pi\nu_0 i}{4}\Tr[D(\nabla\hat{Q})^2+2iz\{\hat{\eps},\hat{\lambda}\} \underline{\delta \hat{Q}}],\no\\
&&+\frac{i}{2}(\pi\nu_0)^2 \langle \Tr[\hat{\lambda}\hat{\theta}^l \sigma^l \underline{\delta\hat{Q}}]\Tr[\hat{\theta}^k\sigma^k\underline{\delta \hat{Q}}]\rangle.\quad
\label{eq:Sd}
\ee
Here, $\hat{Q}$, $\delta\hat{Q}$, $\hat{\lambda}$ and $\hat{\theta}$ are matrices in Keldysh and spin space as well as in the frequency domain. $\Tr$ covers all degrees of freedom including spin as well as an integration over coordinates. The matrix reads $(\hat{\lambda}_{\bfr})_{\eps\eps'}=\hat{\lambda}_{\bfr,\eps-\eps'}$ and the same for $\hat{\theta}$, while $\hat{Q}_{\eps\eps'}$ generally depends on both frequency arguments. The fluctuations of the matrix $\underline{\delta\hat{Q}}$ around its equilibrium (saddle-point) position describe diffusion modes. The structure of $\underline{\delta\hat{Q}}$ will be specified in the next paragraph. Note that the diffusion coefficient term in $S_{dm}$ does not contain the gravitational potentials. The parameter $z$ in the frequency term anticipates its renormalization in the presence of the electron-electron interaction [\onlinecite{Finkelstein83}]; the initial value is $z=1$. The (retarded) diffusion propagator determined by the the first two terms in the action $S_{dm}$ is described by $\mathcal{D}(\bfq,\omega)=1/({D\bfq^2-iz\omega})$, the so-called diffuson. The charge $z$ can be considered as the effective "density of states" of the diffusion modes [\onlinecite{Finkelstein83a,Finkelstein84a, Punnoose05}]. It plays a central role for thermal transport [\onlinecite{Castellani87}], because in the disordered electron liquid $z$ determines the renormalization of the specific heat [\onlinecite{Castellani86}]. As a result of renormalizations, the specific heat becomes $c=zc_{FL}$, where $c_{FL}=2\pi^2\nu_0 T/3$.

The non-interacting saddle point of the matrix $\underline{\hat{Q}}(\bfr,\eps,\eps')$ in $S_{dm}$ is given by the matrix $\hat{\Lambda}$ defined in Keldysh space as
\be
\hat{\Lambda}_{\eps}=\left(\ba{cc} 1&2\mathcal{F}_\eps\\0&-1\ea\right)=\hat{u}_\eps\hat{\sigma}_3\hat{u}_\eps,\quad \hat{u}_\eps=\left(\ba{cc}1&\mathcal{F}_\eps\\0&-1\ea\right),
\ee
where $\mathcal{F}_\eps=\tanh\left(\frac{\eps}{2T}\right)$
is the fermionic equilibrium distribution function; note that $\hat{u}_\eps=\hat{u}^{-1}_\eps$. The deviations from the saddle point are denoted as $\underline{\delta \hat{Q}}(\eps\eps')=u_{\eps}\delta\hat{Q}_{\eps\eps'}u_{\eps'}$, where $\delta \hat{Q}=\hat{Q}-\hat{\sigma}_3$. The physics of disordered systems at low temperatures is dominated by slow diffusive motion of electrons with energies $\lesssim1/\tau$. The diffusion modes can be accounted for by considering gapless fluctuations around the saddle point solution that respect the condition $(\hat{Q}\circ\hat{Q})_{t,t'}=\delta(t-t')$. Here, the $\circ$-symbol denotes a convolution in time. Restricting ourselves to this manifold, a convenient parametrization is given as
\be
\underline{\hat{Q}}=\hat{u}\circ \hat{Q}\circ \hat{u},\quad  \hat{Q}=\hat{U}\circ\hat{\sigma}_3\circ \hat{\overline{U}}.
\ee
Here, $\hat{U}=\hat{U}_{t,t'}(\bfr)$, and $(\hat{U}\circ \hat{\overline{U}})_{t,t'}=\delta(t-t')$.

Finally, the brackets in the last term of the action $S_{dm}$ symbolize the contractions $\langle \theta^0_{k,\bfr,\omega}\theta^{0}_{l,\bfr',-\omega'}\rangle=
\frac{i}{2\nu_0}[\tilde{\Gamma}_{\rho}(\bfr-\bfr')/2]\gamma_2^{kl}2\pi\delta_{\omega-\omega'}$,
and for spin degrees of freedom $\langle \theta^\alpha_{k,\bfr,\omega}\theta^{\beta}_{l,\bfr',-\omega'}\rangle=\frac{i}{2\nu_0}
[\Gamma_{\sigma}\delta_{\bfr-\bfr'}/2]\gamma_2^{kl}2\pi\delta_{\omega-\omega'}\delta_{\alpha\beta}$ with $\alpha,\beta\in \{1,2,3\}$, where
the interaction amplitudes in the singlet and triplet channels, $\tilde{\Gamma}_\rho$ and $\Gamma_\sigma$ respectively, acquire the form
\be
\tilde{\Gamma}_\rho(\bfq)&=&\frac{2\nu_0 V_0(\bfq)+F_0^\rho}{1+(2\nu_0 V_0(\bfq)+F_0^\rho)},\quad
\Gamma_\sigma=\frac{F_0^\sigma}{1+F_0^\sigma}.
\ee
For future purposes it will be convenient to decompose the interaction in the singlet channel into two parts [\onlinecite{Finkelstein83,Finkelstein90}]. One of them is the statically screened Coulomb interaction $\Gamma_0(\bfq)$, while the other one is the short-range interaction $\Gamma_\rho$ which acts within the polarization operator along with $\Gamma_\sigma$,
\be
\tilde{\Gamma}_\rho(\bfq)&=&2\Gamma_0(\bfq)+\Gamma_\rho,\label{eq:tildeG}
\ee
where
\be
\Gamma_0(\bfq)=\frac{\nu_0}{(1+F_0^\rho)^2}\frac{1}{V^{-1}_0(\bfq)+\frac{\partial n}{\partial \mu}},\quad\Gamma_\rho=\frac{F_0^\rho}{1+F_0^\rho}.
\ee
Note that the amplitudes $\Gamma_\rho$ and $\Gamma_\sigma$ acquired a form familiar from Fermi liquid theory. We also obtained the FL-renormalization for $\frac{\partial n}{\partial\mu}$, the quantity which determines the value of the polarization operator in the static limit
\be
\frac{\partial n}{\partial\mu}=\frac{2\nu_0}{1+F_0^\rho}.
\ee

\section{RG analysis of the NL${\bm \sigma}$M in the presence of the gravitational potentials}
\label{sec:RGanalysis}
For the discussion of the dynamical part of the correlation function it is sufficient to expand $\hat{\lambda}\approx 1-\hat{\eta}$ in the action. It will be preferable to use the interaction amplitudes in the form $\frac{1}{2}(\Gamma _{\rho }\delta _{\alpha \delta }\delta _{\beta \gamma }+\Gamma _{\sigma }\bfsigma_{\alpha \delta }\bfsigma_{\beta \gamma })=\Gamma _{1}\delta _{\alpha \delta }\delta _{\beta \gamma }-\Gamma _{2}\delta _{\alpha \gamma }\delta _{\beta \delta }$, where
\be
\Gamma _{1}=\frac{1}{2}(\Gamma _{\rho }-\Gamma _{\sigma }),\quad \Gamma_2=-\Gamma_{\sigma}.
\ee
 To this end, one should consider the following action
\be
S_{\zeta}=\frac{\pi\nu_0 i}{4} \Tr[D(1+\underline{\hat{\zeta}_D})
(\nabla \hat{Q})^2+2iz \{\hat{\eps},1+\underline{\hat{\zeta}_z} \}\delta\hat{Q}]\no\\
+\frac{i}{2}(\pi\nu_0)^2 \sum_{n=0}^2\;\langle \Tr[(1+\underline{\hat{\zeta}_{\Gamma_n}})\underline{\hat{\phi}_n}\delta\hat{Q}]\Tr[\underline{\hat{\phi}_n}\delta\hat{Q}]\rangle,\quad\label{eq:Szeta}
\ee
where $\underline{\hat{\zeta}_{X}}(\bfr,\eps+\omega,\eps)=\hat{u}_{\eps+\omega}\hat{\gamma}_1\hat{u}_{\eps} \zeta_{X}(\bfr,\omega)$ for $X\in\{D,z,\Gamma_0,\Gamma_1,\Gamma_2\}$. In the following we shall also use notations $\zeta_i$ and $X_i$ with $i=1...5$. The contractions for the fields $\phi_n$, $n=0,1,2$  generate the proper interaction terms with $\Gamma_0,\Gamma_1$ and $\Gamma_2$:
\be
&&\langle \phi_0^i(x)\phi_0^j(x')\rangle=\frac{i}{2\nu}\Gamma_0(\bfr-\bfr')\delta(t-t')\gamma_2^{ij},\label{eq:phi0phi0}\\
&&\langle \phi^i_1(x)\phi_1^j(x')\rangle=\frac{i}{2\nu}\Gamma_1\delta(x-x')\gamma_2^{ij},\label{eq:phi1phi1}\\
&&\langle \phi^i_{2,\alpha\beta}(x)\phi_{2,\gamma\delta}^j(x')\rangle=-\frac{i}{2\nu}\Gamma_2\delta_{\alpha\delta}\delta_{\beta\gamma}\delta(x-x')\gamma_2^{ij}.\qquad\label{eq:phi2phi2}
\ee
The initial conditions are obtained from a comparison with Eq.~\eqref{eq:Sd},
\be
\quad\zeta_z=\zeta_{\Gamma_0}=\zeta_{\Gamma_1}=\zeta_{\Gamma_2}=-\eta_1=-\eta_2,\quad \zeta_D=0.\label{eq:init}
\ee
The field $\zeta_D$ was introduced to account for the possibility that the sources migrate to the kinetic term during the RG procedure. The reason why we separate $\zeta_{\Gamma_0}\Gamma_0$ from $\zeta_{\Gamma_1}\Gamma_1$, which are both acting in the singlet channel, is that all the RG-corrections to this channel go to the short-range part only, while the long-range part remains unchanged.

The general structure of the RG-corrections is determined by the number of independent momentum-integrations. Each integration leads to an additional power in the inverse dimensionless conductance, the small parameter of the RG-expansion. At a given order of the RG-expansion, the dependence on the interaction amplitudes can be accounted for to all orders [\onlinecite{Finkelstein83,Finkelstein10}]. It is therefore sufficient to extract the $\zeta_X$-terms from the established RG-diagrams. The procedure is described in detail in Ref. [\onlinecite{Schwiete14b}] for a system with short-range interactions, i.e., in the absence of $\Gamma_0$ and $\xi_{\Gamma_0}$. Here, we report the extension of this analysis to the disordered electron liquid, i.e., to a system with long-range Coulomb interaction.

The final result of the RG analysis based on the extended NL$\sigma$M acquires a very compact form:
\be
\Delta (X_i\zeta_i)=\sum_{j=1}^5 \zeta_j X_j\frac{\partial}{\partial X_j} (\Delta X_i),\label{eq:compactr}
\ee
where $\Delta X$ symbolizes a logarithmic correction to $X$. The result holds for all $X\in\{D,z,\Gamma_0,\Gamma_1,\Gamma_2\}$.

A comment is in order here with respect to Eq.~\eqref{eq:compactr}. There is an important relation connecting different RG-charges [\onlinecite{Finkelstein83,Finkelstein10}],
\be
z-2\Gamma_1+\Gamma_2=\frac{1}{1+F_0^\rho}.\label{eq:relation}
\ee
This relation considerably simplifies the RG-equations. It is worth noting that in Eq.~\eqref{eq:compactr} one has to perform the differentiation first, and use relation \eqref{eq:relation} only afterwards.

It has been explained in Refs. [\onlinecite{Schwiete14a, Schwiete14b}], using the known structure of the RG-equations for the charges $X_i$, that the initial values for the sources \emph{do not change} as a result of renormalization, $\it{provided}$ that initially (i) $\zeta_D=0$ holds and (ii) all remaining $\zeta_Y$ are equal. It turns out that this statement also holds true for the disordered electron liquid considered here and, moreover, the conditions (i) and (ii) are precisely met by the initial conditions stated below Eq.~\eqref{eq:phi2phi2}. As a result, the relations~\eqref{eq:init} are unchanged during the course of the RG. Note the important fact that $\zeta_D$ cannot be generated by the other sources. Thus, we obtained a fixed point in the multi-parametric RG flow, which is a rather non-trivial result for a multi-parametric flow.
Note the multiplicative structure of $S_\zeta$ in this connection: With the exception of $\zeta_D$, the source terms flow \emph{exactly} as their host terms during the course of the RG,
\be
\Delta (X_i\zeta_i)=\zeta_i\Delta X_i,\quad i\in\{2,3,4,5\}
\ee

Finally, let us recall that the RG procedure covers the interval of energies with the elastic scattering rate $1/\tau$ as the upper cutoff and the temperature $T$ as the lower one ($T\ll 1/\tau$).

\section{The RG analysis for the heat density-heat density correlation function ${\bm \chi_{kk}}$}
\label{sec:RGanalysis2}
At this stage, we assume that the RG procedure for the extended NL$\sigma$M has already been completed. For the sake of the presentation, we will temporarily ignore the additional corrections arising from the sub-temperature range in this section.

As usual, we decompose $\chi_{kk}$ into a static and a dynamical part. It is known that the static part is connected with the specific heat renormalized by the electron-electron interactions. As a result of renormalizations it acquires the factor $z$, $\chi_{kk}^{st}=-Tzc_{FL}$ [\onlinecite{Castellani86,Castellani87}]. We now turn to the dynamical part of the correlation function. Since the RG procedure has already been completed, it is permissible to calculate the correlation function in the ladder approximation. In the ladder approximation, only those contributions are selected, which do not include an integration over momenta and frequencies of the diffusion modes.
(This was the purpose of the RG procedure.) For the heat density-heat density correlation function, the relevant contribution originates only from those sources entering the frequency part. We write the corresponding term of the action as
\be
S_{\eps\eta Q}=\frac{1}{2}\pi\nu \gamma^z_{\triangleleft} \Tr[\{\hat{\eps},\hat{\eta}\}\underline{\delta Q}].\label{eq:SetaepsQ}
\ee
Here, the dimensionless coefficient $\gamma^z_{\triangleleft}$ has been introduced in order to account for RG-corrections to the frequency vertex. Then, one obtains
\begin{align}
\chi_{kk}^{dyn}(x_1,x_2)=&2i(\pi\nu)^2 \gamma_{\triangleleft}(\eta_1)\gamma_{\triangleleft}(\eta_2)\int_{\eps\eps'\omega\omega'}\eps\eps'\Delta_{\eps',\omega'}\no\\
&\times \left \langle d^{cl}_{0;\eps_1\eps_2}(\bfr_1)d^q_{0;\eps_2'\eps'_1}(\bfr_2)\right\rangle \mbox{e}^{-it_1\omega+it_2\omega'}, \label{eq:dynpartrescat}
\end{align}
where $\eps_{1,2}=\eps\pm\frac{\omega}{2}$, $\eps'_{1,2}=\eps'\pm\frac{\omega'}{2}$ and the matrices $d^{cl/q}$ describe deviations of $\delta Q$ from the saddle point. Here, we introduced the window function
\be
\Delta_{\eps',\omega'}=\mathcal{F}_{\eps'+\omega'/2}-\mathcal{F}_{\eps'-\omega'/2}.
\ee
The appearance of the window function is characteristic for the dynamical part of the correlation function. For $T\rightarrow 0$, it allows frequencies $\eps'$ to lie in the interval $(\eps'-\omega'/2,\eps'+\omega'/2)$; at finite temperature this range broadens. Still, upon integration in $\eps'$, the function $\Delta_{\eps',\omega'}$ gives rise to a factor of $\omega'$. The correlation function $\left \langle d^{cl}_{0;\eps_1\eps_2}(\bfr_1)d^q_{0;\eps_2'\eps'_1}(\bfr_2)\right\rangle$ describes the diffusive propagation of particle-hole pairs in the singlet channel (only the singlet channel contributes):
\be
&&\left\langle d^{0}_{cl;\eps_1\eps_2}(\bfq) d^0_{q;\eps_3\eps_4}(-\bfq)\right\rangle=-\frac{1}{\pi\nu}\mathcal{D}(\bfq,\omega)\times\label{eq:dsinglet} \\
&&\left(\delta_{\eps_1,\eps_4}\delta_{\eps_2,\eps_3}-
\delta_{\omega,\eps_4-\eps_3}i\pi\Delta_{\eps_1\eps_2}
\tilde{\Gamma}_\rho(\bfq)\tilde{\mathcal{D}}_1(\bfq,\omega)\right).\no
\ee
The second term takes into consideration rescattering induced by the interaction amplitude $\tilde{\Gamma}_\rho$. Here, we introduced the modified diffusion propagator $\tilde{\mathcal{D}}_1(\bfq,\omega)=(D\bfq^2-i\tilde{z}_{1}\omega)^{-1}$ with $\tilde{z}_1(\bfq)=z-\tilde{\Gamma}_\rho(\bfq)$. However, for the expression of $\chi_{kk}^{dyn}(x_1,x_2)$ as given in Eq.~\eqref{eq:dynpartrescat}, rescattering is not effective since it gives rise to an independent frequency integration of the type $\int_\eps \eps \Delta_{\eps+\frac{\omega}{2},\eps-\frac{\omega}{2}}=0$. Without rescattering, one immediately finds
\begin{align}
\chi_{kk}^{dyn}(\bfq,\omega)=&-2\pi\nu \gamma^z_{\triangleleft}(\eta_1)\gamma^z_{\triangleleft}(\eta_2) i \mathcal{D}(\bfq,\omega)\int_{\eps}\eps^2\Delta_{\eps,\omega}\no\\%(\mathcal{F}_{\eps+\frac{\omega}{2}}-\mathcal{F}_{\eps-\frac{\omega}{2}})\no\\
\approx&\gamma^z_{\triangleleft}(\eta_1)\gamma^z_{\triangleleft}(\eta_2)c_0 T\frac{-i  \omega}{D\bfq^2-iz \omega},\label{eq:dynpart}
\end{align}
where the relation $\int_\eps \eps^2\partial_\eps \mathcal{F}_\eps=\pi T^2/3$ has been used.

As we have argued in Refs. [\onlinecite{Schwiete14a, Schwiete14b}], $\gamma^z_{\triangleleft}(\eta_1)=\gamma^z_{\triangleleft}(\eta_2)=z$. Then, after all renormalizations, the dynamical part $\chi_{kk}^{dyn}$ can be written as
\be
\chi_{kk}^{dyn}(\bfq,\omega)=-c_{FL}Tz\frac{iz\omega}{D\bfq^2-iz\omega}.\label{eq:chidyn}
\ee
When joint with the static part, we obtain the full correlation function [\onlinecite{Castellani87}] in a form that is typical for a correlation function of a conserved quantity
\be
\chi_{kk}(\bfq,\omega)=-Tc\frac{D_k\bfq^2}{D_k\bfq^2-i\omega},\label{eq:chifull}
\ee
where $D_k=D/z$ is the heat diffusion coefficient.

It follows from Eq.~\eqref{eq:continuity} for the thermal conductivity that $\kappa=cD_k=c_{FL}D$. We observe an important cancelation between the renormalizations of the heat diffusion coefficient $D_k$ and the specific heat. In combination with the RG results for the conductivity of the disordered Fermi liquid, $\sigma=2e^2\nu_0 D$, this yields the WFL [\onlinecite{Wiedemann1853}]: $\kappa/\sigma=\pi^2 T/3e^2$.
Thus, for  the \emph{virtual transitions} within the RG-interval of energies
described by the standard RG procedure, the WFL holds even in the presence of quantum corrections caused by the interplay of diffusion modes and the electron-electron interaction.

The result is by no means trivial: The WFL should not be considered as a strict law outside the realm of single-particle physics as is already evident from the fact that the electric potential couples to the particle density only, while the gravitational potential probes the entire Hamiltonian density. Still, the WFL holds for the leading logarithmic corrections arising from RG processes in a two-dimensional disordered electron system.

A perturbative analysis of logarithmic corrections to the heat density-heat density correlation function in a two-dimensional electron gas, i.e., in a system with long-range Coulomb interaction has been performed in Refs. [\onlinecite{Castellani87,Schwiete15}]. The corrections from the RG-interval can be used to formulate the RG equations, according to the usual procedure [\onlinecite{Castellani84}]. The use of the extended NL$\sigma$M, in turn, allows to perform the RG analysis in a more systematic fashion.

\section{The heat density-heat density correlation function $\chi_{kk}$ in the presence of the sub-thermal corrections}
\label{sec:sub}

In contrast to the RG integrals, the sub-thermal contributions are determined by the combination $(\mathcal{F}_{\eps+\nu}+\mathcal{F}_{\eps-\nu})$, where $\nu$ is the frequency transferred by the electron interaction. Due to this combination, the transferred frequency is limited either by temperature or by the electron frequency $|\eps|$. The effect crucially depends on the dynamically screened Coulomb interaction, $\tilde{\Gamma}_{0;d}^R(\bfk,\nu)$, for which one has the well known expression:
\be
\tilde{\Gamma}^R_{0;d}(\bfk,\nu)=\frac{\nu_0}{(1+F_0^\rho)^2}\frac{1}{V^{-1}_0(\bfk)+\frac{\partial n}{\partial \mu}\frac{D_{FL}\bfk^2}{D_{FL}\bfk^2-i\nu}},\label{eq:Gamma0ds}
\ee
where the electron density diffusion coefficient is $D_{FL}=(1+F_0^\rho)D$. $\tilde{\Gamma}^R_{0;d}$ describes the dynamically screened Coulomb interaction dressed by short-range amplitudes both at the external vertices and within the polarization operator.

For the sub-thermal corrections, the momentum integration is determined by small momenta. For a given frequency $\nu$, most important momenta fulfill the inequality $|\nu|/(D\kappa_s)<k<\sqrt{|\nu|/D}$, where $\kappa_s=4\pi e^2 \nu_0$ is the inverse screening radius. In this interval, one can approximate the dynamically screened interaction as
\be
\tilde{\Gamma}_{0;d}^R(\bfk,\nu)\approx -\frac{1}{2(1+F_0^\rho)^2}\frac{i\nu}{D\bfk^2}.\label{eq:imaginV}
\ee
Eventually, the bare $1/D\bfk^2$ singularity gives rise to logarithmic corrections. It is now clear that the contributions from the sub-temperature interval are important in the case of the dynamically screened Coulomb interaction, for which $\tilde{\Gamma}_{0;d}^R(\bfk,\nu)$ is singular. For a short-range interaction the discussed interval of frequencies does not exhibit any singularity and, therefore, is not important.

Note that the interval $|\nu|/(D\kappa_s)<k<\sqrt{|\nu|/D}$ is also responsible for the double-logarithmic dependence of the tunneling density of states as well as other spurious corrections that appear in intermediate stages of the RG procedure. In the case of the sub-thermal term only a single-logarithm arises, because allowed frequencies $\nu$ are small, of the order of the temperature. In contrast, a double-logarithmic dependence is obtained for those integrals, for which the frequency takes large values exceeding $T$.

In our recent paper [\onlinecite{Schwiete15}], we have checked that on a perturbative level Eqs.~\eqref{eq:chidyn} and ~\eqref{eq:chifull} preserve their form even in the presence of the sub-thermal corrections. The only modification is the diffusion coefficient
\be
\tilde D=D+\delta D.
\ee
We now generalize this result by including the renormalized parameters from the RG interval. Since $\delta D$ is a correction, one can expand $\chi_{kk}^{dyn}$. The result can be written as
\be
\delta\chi_{kk}^{dyn}(\bfq,\omega)=c_{FL}Tiz^2\omega (\delta D)\bf q^2{\mathcal D}^2(\bfq,\omega).\label{eq:chicorr}
\ee

In a diagrammatic language, the corrections to $\delta\chi_{kk}^{dyn}$ which determine $\delta D$ originate from two contributions. One of the two contributions is found from diagrams with a single vertical interaction line inside a diffusion ladder. (Such corrections to the diffusion coefficient arise in the sub-thermal regime only.) The other contribution to $\delta D$ originates from drag diagrams, which contain a product of two electron interaction amplitudes.

%originates from the diagrams with a single vertical interaction line inside a diffusion propagator. Such a correction to the diffusion coefficient can be only from the sub-thermal terms.
First, we consider those changes to the heat density-heat density correlation function from diagrams with vertical interaction lines which have their origin in a correction to $\delta D$:
\begin{align}
\delta\chi_{kk}^{dyn}(\bfq,\omega)&=2i\pi z^2D{\bfq}^2\mathcal{D}_{\bfq,\omega}^2\int_{\eps}\eps\Delta_{\eps,\omega} \label{eq:chi2kk}
\\
&\times \int_{\bfk,\nu} \nu^2 (\mathcal{F}_{\eps-\nu}+\mathcal{F}_{\eps+\nu})\mathcal{D}_{\bfk,\nu}^2 \mbox{Im}\Gamma_{0;d}^R (\bfk,\nu).\no
\end{align}
Here, the amplitude $\Gamma_{0;d}^{R}$ is defined as
\be
\Gamma_{0;d}^R=\tilde{\Gamma}_{0;d}^R\frac{\mathcal{D}_1^2}{\mathcal{D}^2},\qquad \mathcal{D}_1(\bfk,\nu)=\frac{1}{D\bfk^2-\frac{i\nu}{1+F_0^\rho}}.
\ee
%and $\Gamma^A_{0;d}(\bfk,\nu)=\Gamma^R_{0;d}(\bfk,-\nu)$
It includes the dynamical dressing of external vertices of the screened Coulomb interaction caused by the short-range amplitudes. In the interval of small momenta indicated above Eq.~\eqref{eq:imaginV} one may approximate
\be
\Gamma_{0;d}^R (\bfk,\nu)=\frac{1}{2(1+F_0^\rho)}\frac{\mathcal{D}_1(\bfk,\nu)}{D{\bfk}^2}\frac{1}{\mathcal{D}^2(\bfk,\nu)}.
\ee

%where in this formula $\mathcal{D}=[D(\bfk+\bfq/2)^2-iz\nu]^{-1}$ and $\mathcal{D}_1=[D(\bfk+\bfq/2)^2-i{\nu}/{(1+F_0^\rho)}]^{-1}$.
The full result for the drag contribution to the heat density-heat density correlation function reads:
\begin{align}
&\delta\chi_{kk}^{dyn}(\bfq,\omega)=-2i\pi z^2\;\mathcal{D}_{\bfq,\omega}^2\int_{\eps}\eps\;\Delta_{\eps,\omega}\label{eq:chi3kk}\\
&\times \int_{\bfk,\nu} \nu^2 (\mathcal{F}_{\eps-\nu}+\mathcal{F}_{\eps+\nu})\Gamma_{0;d}^R (\bfk+\bfq/2,\nu)\Gamma_{0;d}^A (\bfk-\bfq/2,\nu)\no\\
&\qquad\qquad \times\mathcal{D}_{\bfk+\bfq/2,\nu}(\mathcal{D}_{\bfk+\bfq/2,\nu}+\overline{\mathcal{D}}_{\bfk-\bfq/2,\nu})|_{q^{2}}.\no
\end{align}
Here, the notation $|_{q^{2}}$ indicates that the contribution proportional to $\bf{q}^{2}$ needs to be extracted from the integral in order to get $\delta D$. The advanced diffuson $\overline{\mathcal{D}}$ is related to the retarded diffusion as $\overline{\mathcal{D}}_{\bfq,\omega}=\mathcal{D}_{\bfq,-\omega}$; the amplitude $\Gamma_{0;d}^{A}$ is defined as
%\be
%\Gamma_{0;d}^R=\tilde{\Gamma}_{0;d}^R\frac{\mathcal{D}_1^2}{\mathcal{D}^2},\qquad %\mathcal{D}_1(\bfq,\omega)=\frac{1}{D\bfq^2-\frac{i\omega}{1+F_0^\rho}}
%\ee
$\Gamma^A_{0;d}(\bfk,\nu)=\Gamma^R_{0;d}(\bfk,-\nu)$.
%include the dynamical dressing of the external vertices of the screened Coulomb interaction caused by the short-range amplitudes. In the interval of small momenta indicated above Eq.~\eqref{eq:imaginV},
%\be
%\Gamma_{0;d}^R (\bfk+\bfq/2,\nu)=\frac{1}{2(1+F_0^\rho)}\frac{\mathcal{D}_1}{D(\bfk+\bfq/2)^2}\frac{1}{\mathcal{D}^2},
%\ee
%where in this formula $\mathcal{D}=[D(\bfk+\bfq/2)^2-iz\nu]^{-1}$ and $\mathcal{D}_1=[D(\bfk+\bfq/2)^2-i{\nu}/{(1+F_0^\rho)}]^{-1}$.

The correction to $\delta D$ arising from the drag diagrams is twice smaller than the one from the vertical diagrams and, furthermore, they are of opposite signs. Correspondingly, the "incoming" contribution described by the vertical diagram with the dynamically screened Coulomb interaction line dominates. In this context it is worth noting, however, that due to the presence of other diagrams whose contributions cancel among each other, the final answer could as well be attributed to different diagrams. Quite generally, an unambiguous interpretation of individual diagrams in terms of physical processes is not possible.

The total correction to the thermal conductivity from sub-thermal energies reads
\begin{align}
\delta \kappa =\frac{T}{12} \log \frac{D\kappa_s^2}{T}\label{eq:kappadelta}.
\end{align}
This result coincides with the one existing in the literature, Refs.~[\onlinecite{Michaeli09,Raimondi04,Niven05,Catelani05}]. Here, it has been obtained with the use of the heat density-heat density correlation function. This allowed us to check that all Fermi-liquid renormalization and the parameter $z$ as well as all other RG renormalizations drop out when calculating this correction.

\section{Thermal transport near the metal insulator transition in Si-MOSFETs}
\label{sec:MOSFET}
In this section, we apply the developed theory to the description of thermal transport on the metallic side of the metal-insulator transition in Si-MOSFETs. This requires the inclusion of a valley degree of freedom $n_v$ into the theory, as discussed in Ref.~[\onlinecite{Punnoose01}]. The coupled RG equations that govern the flow of the dimensionless resistance $\rho=(4\pi^2\nu n_v D)^{-1}$ in combination with the interaction amplitude $w_2=\Gamma_2/z$ read [\onlinecite{Punnoose01}]
\be
\frac{\partial \rho}{\partial \xi}&=&\rho^2\left[n_v+1-(4n_v^2-1)\right.\no\\
&&\left.\qquad\times\left(\frac{1+w_2}{w_2}\ln(1+w_2)-1\right)\right],\\
\frac{\partial w_2}{\partial \xi}&=&\frac{1}{2}\rho{(1+w_2)^2}.\label{eq:coupled}
\ee
The temperature dependence is encoded in the logarithmic variable $\xi=\log({1}/{T\tau})$. The first term in the square brackets in the equation for $\rho$ originates from the Cooperon degrees of freedom and increases the tendency of the system to localize. It is worth mentioning that the weak localization terms induced by the Cooperons do not violated the WFL.

The solution of the coupled RG equations \eqref{eq:coupled} depends on the number of valleys $n_v$. By contrast, the correction to the thermal conductivity from sub-thermal energies is independent of $n_v$. The argument goes as follows: For the diagrams corresponding to the contributions to $\delta D$ discussed in Sec.~\ref{sec:sub} the number of closed fermionic loops coincides with the number of interaction lines. Each loop is proportional to $n_v$, while the interaction line in the case of the screened long-range Coulomb interaction is inversely proportion to $n_v$. As a result, the $n_v$-factors cancel out. Thus, the finite number of valleys does not change the result given in \eqref{eq:kappadelta}.

\begin{figure}
\includegraphics[width=8cm]{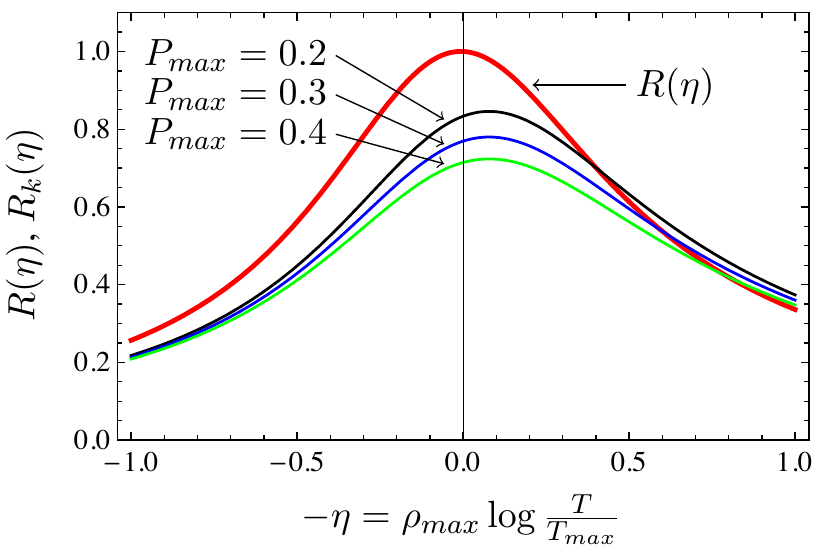}
\caption{"Thermal resistance" $R_k$ alongside with the resistance $R$; both quantities are presented in dimensionless units. $R_k$ coincides with $R$ if the WFL holds. In the discussed theory, the maxima $R_k$ are shifted from the maximum of $R$ by a value which does not depend on the parameter $P_{max}$, which is used as the measure for the violation of the WFL law. }
\label{fig:Rkofeta}
\end{figure}

Here, we limit ourselves to the case $n_v=2$, which corresponds to Si MOSFETs. The dimensionless resistance displays a maximum, $\rho_{max}$, at a temperature denoted as $T_{max}$. At $T_{max}$, the interaction amplitude takes the universal value $w_2=0.457$. It is convenient to introduce the new logarithmic variable $\eta=\rho_{max}\log(T_{max}/T)$. The structure of the RG equations \eqref{eq:coupled} implies that the normalized resistance $R(\eta)=\rho(\eta)/\rho_{max}$ is a universal function of $\eta$. $R(\eta)$ is displayed in Fig.~\ref{fig:Rkofeta}.

Turning to the description of thermal transport, we first define the "thermal resistance" as
\be
\rho_k=\frac{e^2}{2\pi^2}\frac{\mathcal{L}_0 T}{\kappa}
\ee
This definition is motivated by the fact that $\rho_k=\rho$ holds as long as the WFL is satisfied. It is therefore natural to introduce the dimensionless function $R_k(\eta)=\rho_k(\eta)/\rho_{max}$, the thermal analog of $R(\eta)$.

Let us now consider the effect of the WFL violating correction $\delta \kappa$ of Eq.~\eqref{eq:kappadelta} on the function $R_k$. $R_k(\eta)$ can be parametrized as
\be
R_k(\eta)=\frac{R(\eta)}{1+R(\eta)(P_{max}+\frac{1}{2}\eta)},\label{eq:Rk}
\ee
where
\be
P_{max}\equiv\frac{1}{2}\rho_{max}\log\frac{D\kappa_s^2}{T_{max}}=\frac{\rho_{max}-\rho_k(0)}{\rho_k(0)}
\ee
is a measure for the violation of the WFL law at the temperature $T_{max}$.
Note that in Eq.~\eqref{eq:Rk} we neglect the RG-flow of $D=D(\eta)$ which can appear as a refinement of $\eta/2$ in the denominator of $R_k(\eta)$; such sub-leading terms are beyond the accuracy of our RG procedure.

Remarkably, the maximum of $R_k(\eta)$ is positioned at a universal value (independent of $P_{max}$), determined by the condition $R'/R^2=1/2$. This gives $\eta_{k,max}\approx-0.0785$. Examples for curves with $P_{max}=0.2$, $0.3$ and $0.4$ are shown in Fig.~\ref{fig:Rkofeta}.

\section{Conclusion}

In this paper, we developed a consistent theory of thermal transport for the disordered electron liquid, i.e., for a system with long-range Coulomb interactions as well as short range Fermi-liquid type interactions in the presence of disorder. To this end we incorporated Luttinger's gravitational potentials into the Keldysh NL$\sigma$M formalism and presented an analysis of logarithmic corrections to the heat density-heat density correlation function. We used a two-stage procedure: first the RG procedure was implemented within the RG-interval $(1/\tau,T)$. This stage does not reveal a deviation from the WFL. We used the results of the RG analysis as a starting point for the second stage. Here, we considered sub-thermal corrections violating the WFL. These corrections originate from processes with excited electron-hole pairs in the sub-thermal interval of energies. The sign of the overall correction is positive, $\delta \kappa >0$.
%The excess thermal conductivity is described by a diagram which corresponds to a drag process for the heat density. In contrast, the corrections in the diffusive limit lead to an \emph{increase} of the thermal conductivity.
The positive sign %of the correction
indicates that the incoming scattering processes are dominant. Loosely speaking, in the diffusive case with long-range Coulomb interaction, electrons can use the energy $\sim T$ from a remote region to facilitate heat transfer. Based on this theory, we made a prediction for the temperature dependence of the thermal resistance on the metallic side of the metal-insulator transition in Si MOSFETs, as illustrated in Fig.~\ref{fig:Rkofeta}. This is the main result of this paper.

\section*{Acknowledgments}

This paper is our contribution to a special issue dedicated to Prof.~L.~V.~Keldysh's 85th Birthday. The Keldysh technique was absolutely crucial for developing the microscopic theory of thermal conductivity presented in this paper.

The authors thank K. Behnia, M. Brando, C.~Fr\"a\ss dorf, M. Feigel'man, I.~Gornyi, I.~Gruzberg, G.~Kotliar, T.~Kottos, B.~Shapiro, A.~Mirlin, E.~Mishchenko, J. Schmalian, J.~Sinova and C. Strunk for discussions. The authors gratefully acknowledge the support by the Alexander von Humboldt Foundation. The work at the Landau Institute for Theoretical Physics (AF) was supported by the Russian Science Foundation under the grant No. 14-42-00044. The work at Texas A\&M University (AF) is supported by the U.S. Department of Energy, Office of Basic Energy Sciences, Division of Materials Sciences and Engineering under Award DE-SC0014154.

%\bibliography{library-1a,library-2}

\begin{thebibliography}{36}
\bibitem[{Hea()}]{Heat}
\bibinfo{note}{We refer to the `grand canonical energy' $k=h-\mu n$ as `heat
  density' because $dk=Tds$ ($s$ being the entropy density) holds for the
  differential under the condition of a constant chemical potential. We would
  like to stress that this identification is not valid under general
  conditions.}
\end{thebibliography}

\end{document}